\documentclass[aps,prl,reprint,showpacs]{revtex4-1}

\usepackage{graphicx}
\usepackage{subfigure}
\usepackage[pdftitle={A high-rate source for single photons in a pure quantum state},pdfauthor={Christoph Kurz},pdfdisplaydoctitle=true,bookmarks=false,pdflang={en-US}]{hyperref}

\begin{document}

\title{A high-rate source for single photons in a pure quantum state}

\author{Christoph Kurz$^1$}
\email{c.kurz@physik.uni-saarland.de}
\author{Jan Huwer$^{1,2}$}
\author{Michael Schug$^1$}
\author{Philipp M\"uller$^1$}
\author{J\"urgen Eschner$^{1,2}$}
\affiliation{$^1$Universit\"at des Saarlandes, Experimentalphysik, Campus E2 6, 66123 Saarbr\"ucken, Germany\\
             $^2$ICFO -- Institut de Ci\`{e}ncies Fot\`{o}niques, Mediterranean Technology Park, 08860 Castelldefels (Barcelona), Spain}

\date{\today}

\begin{abstract}
We report on the efficient generation of single photons, making use of spontaneous Raman scattering in a single trapped ion. The photons are collected through in-vacuum high-NA objectives. Photon frequency, polarization and temporal shape are controlled through appropriate laser parameters, allowing for photons in a pure quantum state.
\end{abstract}


\maketitle

Quantum networks allow for quantum communication between distant locations (nodes) where local quantum information processing is carried out. The key ingredient for this kind of network architecture is to establish entanglement between network nodes \cite{entanglement}. Its scalability is facilitated by employing quantum repeaters \cite{repeater1,repeater2}.

Different platforms are pursued for implementing quantum networks, and various schemes exist to generate entanglement between their nodes. In the field of trapped single atoms and ions as nodes \cite{nodes}, one approach is to use pairs of entangled photons, split them up and have them absorbed by two separate atoms, thus transferring the photonic entanglement onto the atoms \cite{enttransfer}. Alternatively, each atom is first entangled with a single photon through an emission process \cite{apentanglement1,apentanglement2,apentanglement3}. The photons are then brought to interference on a beam splitter, and a measurement is performed that projects the atoms into an entangled state \cite{entswapping1,entswapping2,entswapping3}. A hybrid version of these two approaches is to have one atom emit a single photon (with which it is entangled) that is absorbed by the other, thereby establishing entanglement between the atoms \cite{photonexchange}. All these schemes are probabilistic, but if there is a suitable process heralding the successful creation of entanglement, subsequent information processing operations can be performed deterministically.

For these schemes to succeed in practice, a high rate of generated and collected single photons in a well-defined quantum state is necessary. Two main ways of achieving this have emerged so far. One is the use of optical cavities strongly coupled to the atomic systems and to optical fibers (operating as quantum channels) \cite{apentanglement3,cavity1,cavity2,cavity3,cavity4}; this allows for efficient exchange of light between atoms and fibers. Alternatively, a free-space approach employing high-NA objectives is pursued \cite{highna1,highna2,highna3,highna4}, trading better fiber coupling for higher rates.

In previous experiments, we implemented building blocks of some of these schemes. Pairs of entangled photons from spontaneous parametric down-conversion were split up and one of the photons of a pair was absorbed by a single ion \cite{heraldedabsorption1,heraldedabsorption2}. Temporal, frequency and polarization correlations between the absorption process and the partner photon showed that this is a promising approach to create entanglement between remote quantum systems. We also demonstrated the generation of single photons tunable in bandwidth and temporal shape \cite{singlephotons}. The individual and mutual coherence properties of photons emitted by single ions in two spatially separated traps was investigated through two-photon interference. However, the generated photons were in a mixed frequency state. Here we report on the efficient generation of single photons in a pure state of both polarization and frequency without the need for optical filtering elements. Moreover, we achieved an increased rate of fiber-coupled photons as well as higher tunability of the photon wave packets compared to our previous work.

\begin{figure}[ht]
   \includegraphics[width=0.9\columnwidth]{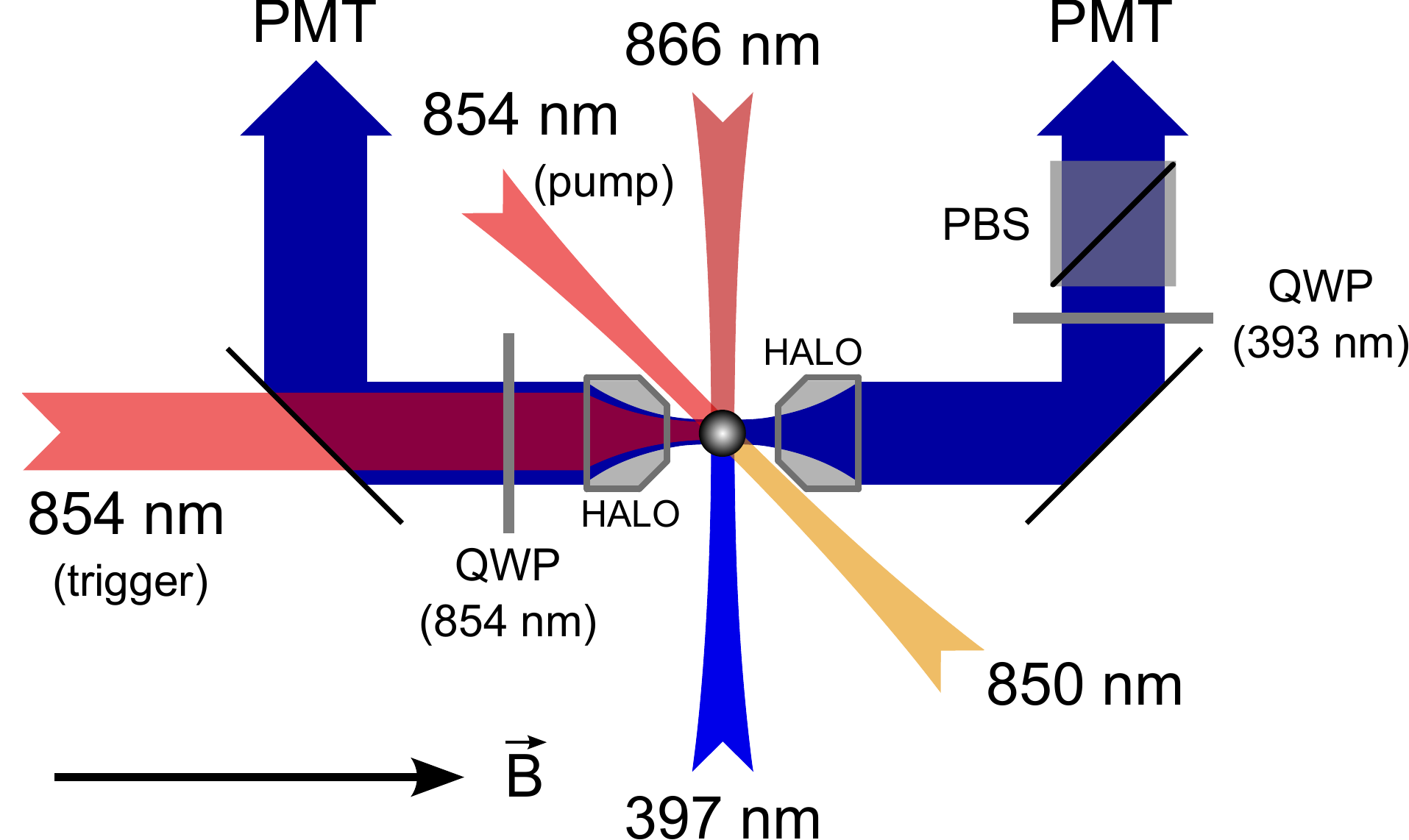}
   \caption{\label{fig:setup} Experimental setup. \mbox{HALO}: high-NA laser objective, PMT: photomultiplier tube, QWP: quarter-wave plate, PBS: polarizing beam splitter, $\vec{\rm B}$: magnetic-field direction. The ion is trapped between the \mbox{HALOs}.}
\end{figure}

Our experimental setup is sketched in fig. \ref{fig:setup}. Single $^{40}$Ca$^+$ ions are trapped in a linear radio-frequency (Paul) trap and Doppler cooled by frequency-stabilized diode lasers at 397~nm and 866~nm wavelength \cite{laserlock}. Additional laser beams at 850~nm and 854~nm are used for state preparation and single-photon triggering. This is facilitated by a magnetic field, also defining a quantization axis. The ion is positioned between two high-NA laser objectives (\mbox{HALOs}, ${\rm NA}=0.4$) for fluorescence collection and laser focusing, each covering $\sim$~4.2\% of the total solid angle \cite{halos}. Blue fluorescence light (either at 397~nm or 393~nm) is coupled to optical fibers and detected by photomultiplier tubes (PMTs) with $\sim$~28\% quantum efficiency. Their output pulses are time-tagged and stored for later processing. Wave plates are used to adjust the polarizations of the laser beams and to analyze the polarization of the fluorescence.

\begin{figure}[ht]
   \includegraphics[width=0.95\columnwidth]{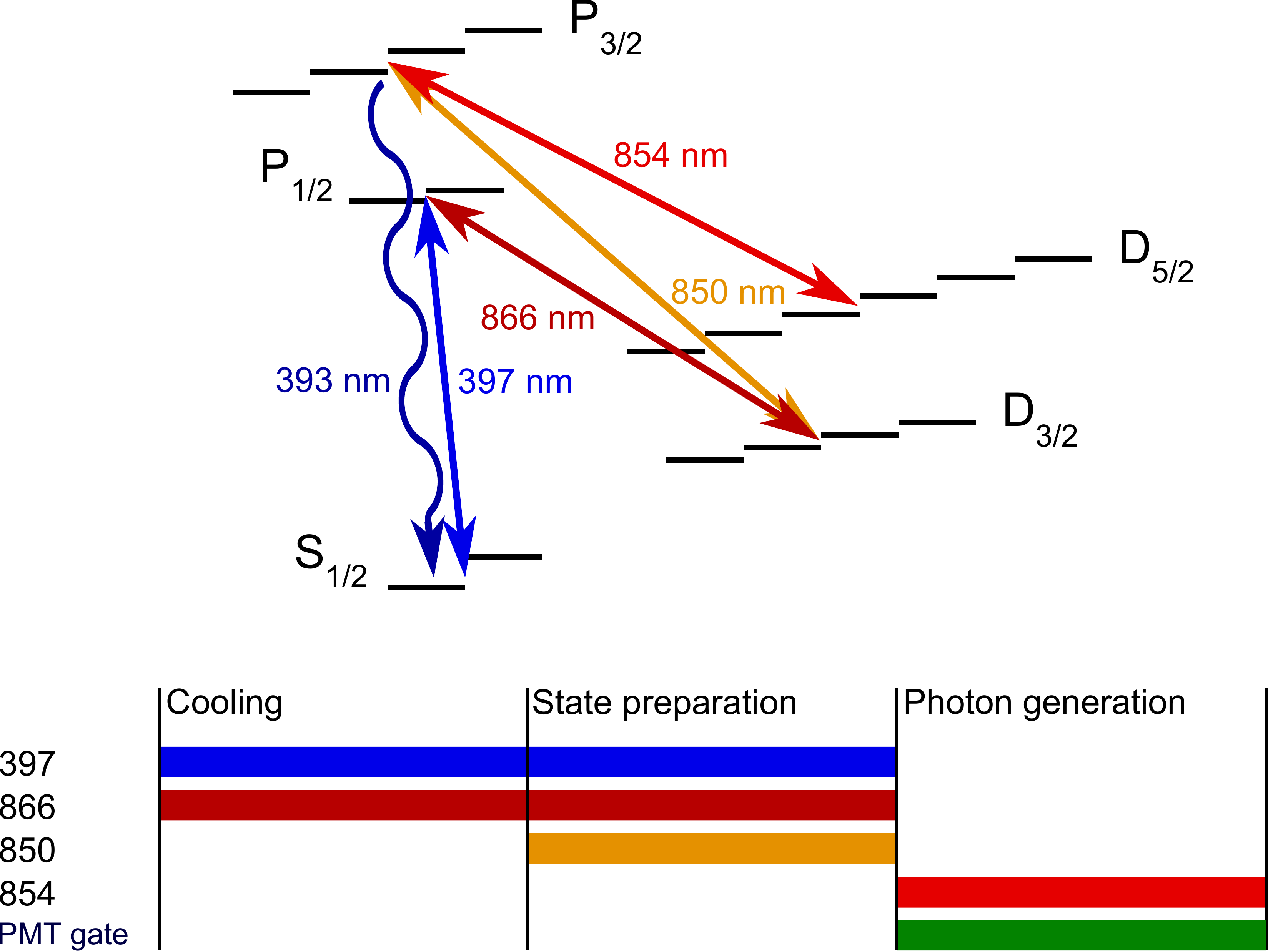}
   \caption{\label{fig:sequence} $^{40}$Ca$^+$ level scheme including Zeeman sub-states and sequence of laser pulses for cooling, state preparation and photon generation.}
\end{figure}

The experimental sequence is outlined in fig. \ref{fig:sequence}. First, the ion is Doppler cooled by the 397~nm and 866~nm lasers for 600~ns. Then the 850~nm laser is switched on for 2~$\mu$s, transferring the ion to the long-lived D$_{5/2}$ state. Here we make use of a three-photon resonance excited by the three lasers, coupling the S$_{1/2}$ ground state directly to the P$_{3/2}$ excited state, i.e. with minimized population of the P$_{1/2}$ and D$_{3/2}$ level. This brings the rate with which population is transferred to the D$_{5/2}$ state close to the limit set by the lifetime and branching ratio of the P$_{3/2}$ state. Subsequently, the 854~nm trigger laser is switched on for a variable time (800~ns -- 35~$\mu$s) while all other lasers are off, driving the D$_{5/2}$--P$_{3/2}$ transition and releasing a single Raman-scattered photon at 393~nm wavelength.

\begin{figure}[ht]
   \subfigure[]{\includegraphics[width=0.9\columnwidth]{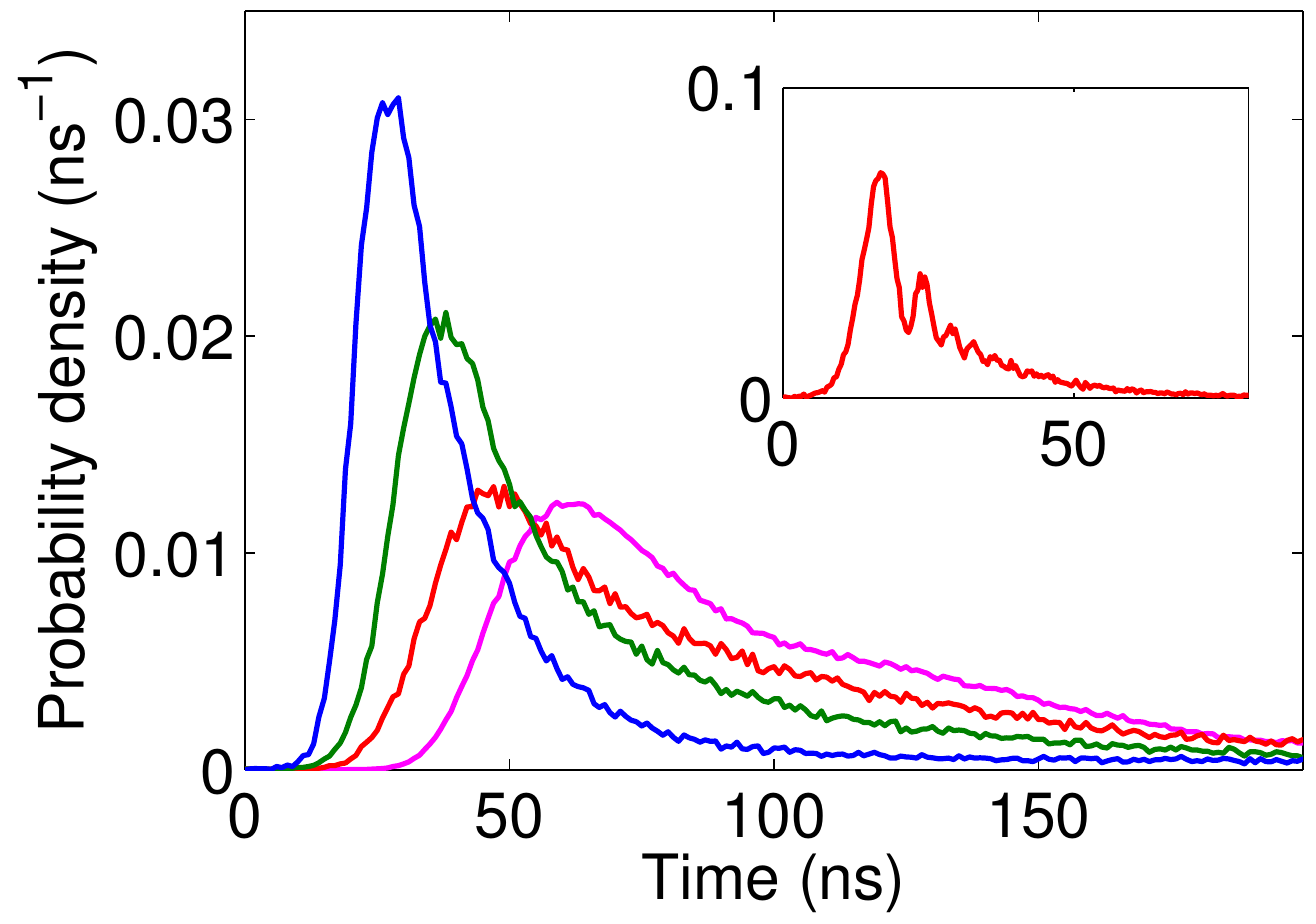} \label{fig:arrivaltimes}}
   \subfigure[]{\includegraphics[width=0.9\columnwidth]{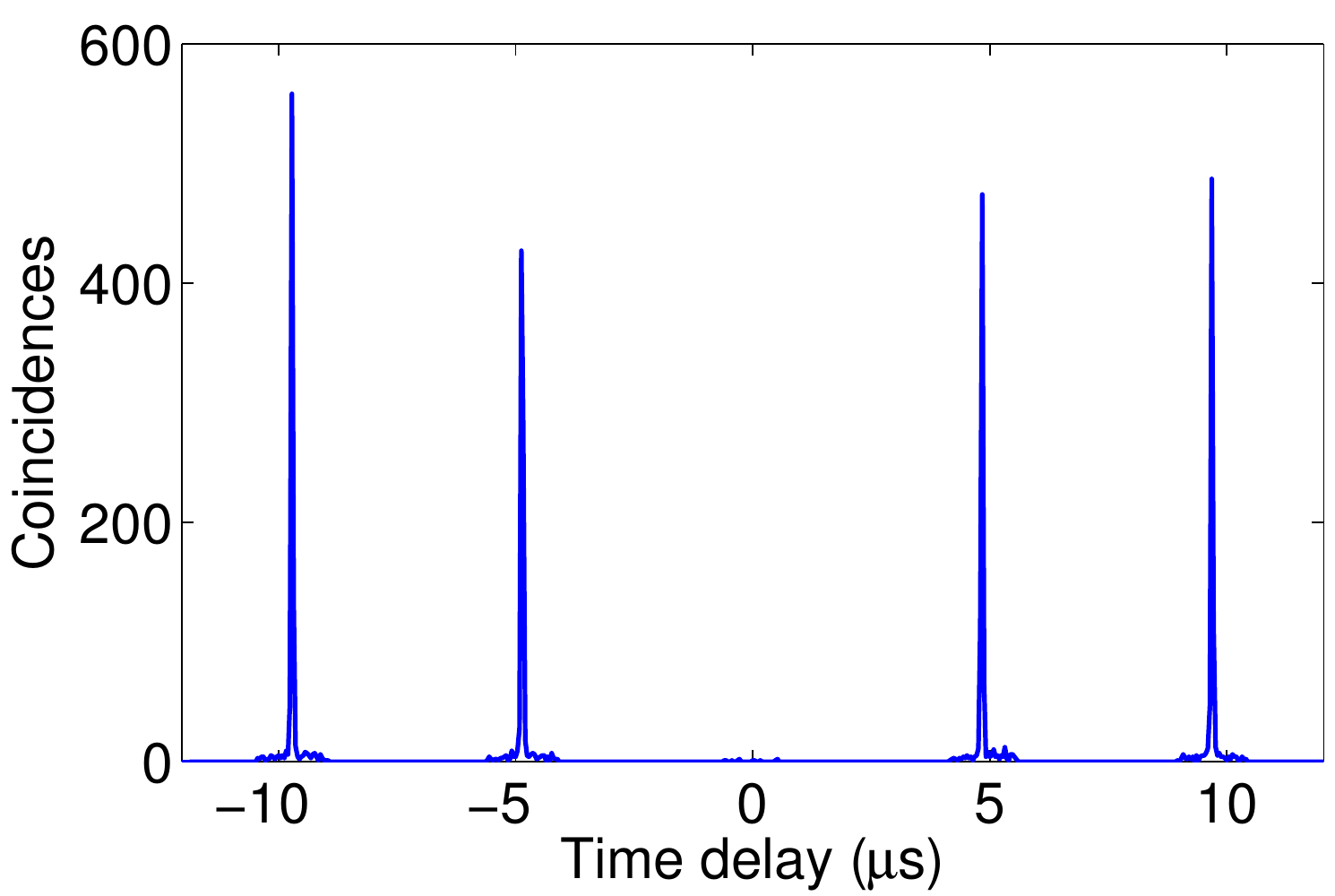} \label{fig:g2}}
   \caption{(a) Arrival time distribution of 393~nm photons. From the shortest to the longest wave packet, the 854~nm laser intensity is $520\frac{{\rm W}}{{\rm cm}^2}$, $67\frac{{\rm W}}{{\rm cm}^2}$, $17\frac{{\rm W}}{{\rm cm}^2}$ and $0.89\frac{{\rm W}}{{\rm cm}^2}$, respectively. Inset: For higher laser powers, Rabi oscillations become visible. (b) Second-order time correlation (g$^{(2)}$) function of the detected single photons (bin size: 40~ns, measuring time: 6~min). The repetition rate for single-photon generation is 206~kHz.}
\end{figure}

For several values of 854~nm laser power, a histogram of arrival times of the single 393~nm photons is displayed in fig. \ref{fig:arrivaltimes}, reflecting control of the temporal shape of the photonic wave packet. The higher the laser power, the faster the D$_{5/2}$ state is depopulated and hence the shorter is the wave packet. The inset shows a lifetime-limited photon of $\sim$~14~ns length, twice as long as the decay time of the P$_{3/2}$ excited state. As the 854~nm laser beam is focused onto the ion through one of the \mbox{HALOs}, an optical power of 3.1~$\mu$W is sufficient. On the other hand, the longest photon created has a duration of 1.03~$\mu$s at 13~nW power of 854~nm.

To assess the single-photon character of the scattered light, we measure the second-order time correlation (g$^{(2)}$) function (fig. \ref{fig:g2}). 393~nm light is collected from both \mbox{HALOs} and detected with individual PMTs. This is equivalent to a Hanbury-Brown-Twiss setup, where the ion itself plays the role of the beam splitter \cite{hbt}. The detected photon arrival times on the two PMTs are then correlated. For a perfect single-photon source, we do not expect any coincidences at zero time delay. From the measured residual counts we deduce a multi-photon to single-photon ratio for light emitted by the ion of 1.34(63)\%, corrected for accidental coincidences caused by stray light and detector dark counts. This small amount of multi-photon events is attributed to the acousto-optic modulators switching the lasers in the state preparation phase. If a single photon is emitted while these lasers are not yet fully switched off, a non-zero probability to re-excite the ion to the P$_{3/2}$ state arises, sometimes resulting in a second photon.

The highest repetition rate in these experiments is 230~kHz, given by optimizing the state preparation time for the highest rate of generated photons. Together with a total detection efficiency of 1.18\% from both \mbox{HALOs} (including the PMT quantum efficiencies and using multi-mode optical fibers), we obtain a rate of fiber-coupled single photons of 5.46~kHz (1.53~kHz detected). For the highest coupling efficiency we could achieve into single-mode fibers, we find a rate of 1.44~kHz (400~Hz detected). The value for the single-mode detection rate compares well with results from cavity (350~Hz \cite{cavity1}) and free-space (1.14~kHz \cite{quantuminterference}) experiments.

In the results shown so far, the ion was prepared in a statistical mixture of all six Zeeman states of the D$_{5/2}$ manifold. Repumping back to the S$_{1/2}$ ground state then happens through a multitude of Raman transitions, resulting in a single photon in a mixed polarization and frequency state. In schemes involving two-photon interference \cite{singlephotons}, this limits the indistinguishability of the two photons. Therefore we implemented an optical pumping scheme that populates only one specific D$_{5/2}$ Zeeman sub-level. This is achieved by using an additional 854~nm pump beam during state preparation, entering the trap at $45^\circ$ with respect to the quantization axis. Its polarization is chosen such that only $\pi$ and $\sigma^+$ transitions are driven, effectively pumping all population to the $\left|{\rm D}_{5/2}{\rm, m}=+\frac{5}{2}\right\rangle$ state. After preparation of this pure state, the 854~nm trigger laser is turned on to drive a Raman transition via the $\left|{\rm P}_{3/2}{\rm, m}=+\frac{3}{2}\right\rangle$ state to the $\left|{\rm S}_{1/2}{\rm, m}=+\frac{1}{2}\right\rangle$ ground state, thereby generating a single 393~nm photon in a pure polarization ($\left|\sigma^+\right\rangle$) and frequency state. By adjusting the laser polarizations appropriately, we are also able to generate single photons in the $\left|\sigma^-\right\rangle$ polarization state.

\begin{figure}[ht]
   \subfigure[]{\includegraphics[width=0.9\columnwidth]{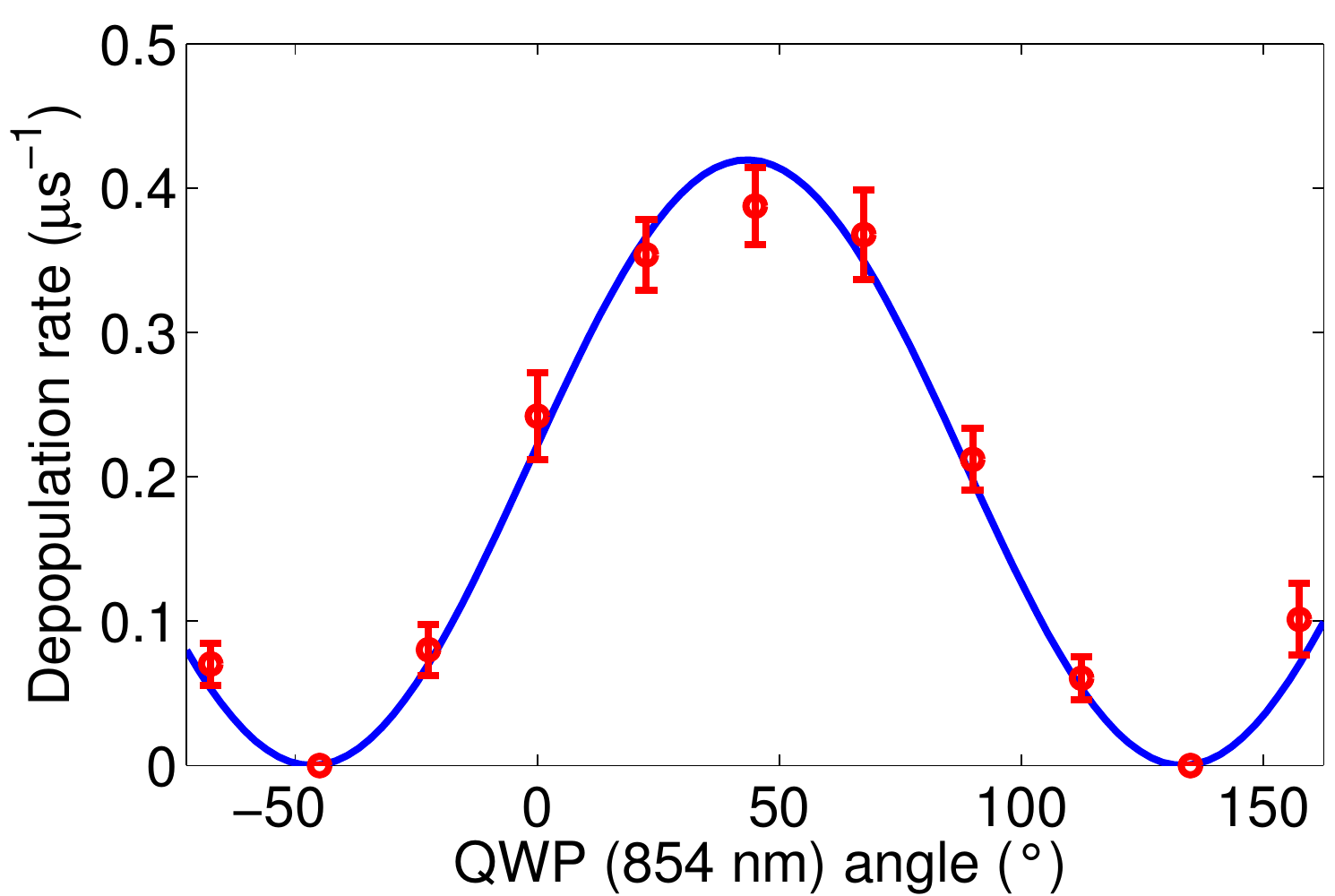} \label{fig:depopulation}}
   \subfigure[]{\includegraphics[width=0.9\columnwidth]{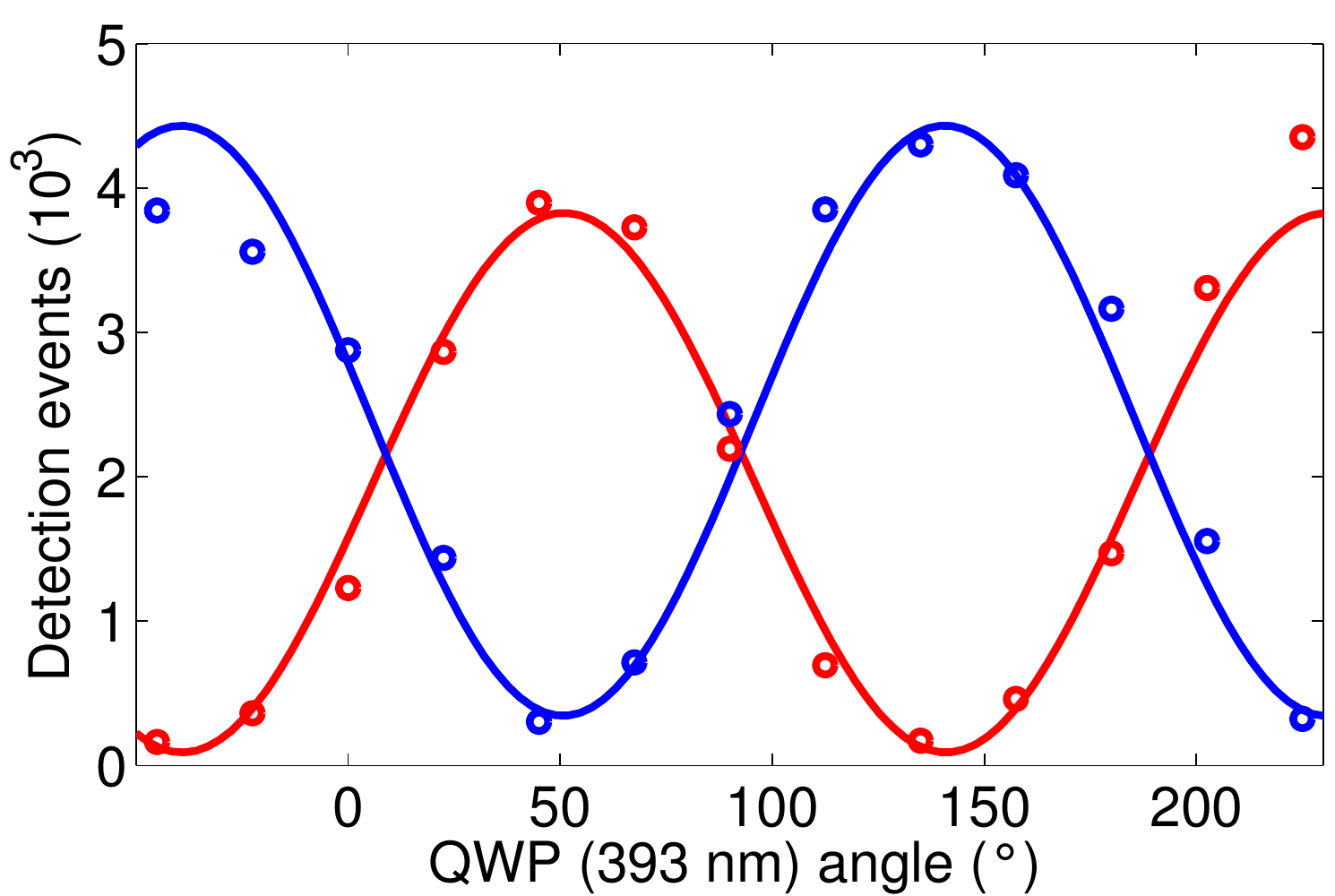} \label{fig:polarization}}
   \caption{(a) Depopulation rate (inverse photon duration) of the D$_{5/2}$ state as a function of the 854~nm quarter-wave-plate angle for non-saturating 854~nm laser power during the photon generation phase. (b) Number of photon detection events as a function of the 393~nm quarter-wave-plate angle for two different D$_{5/2}$ Zeeman states ($\left|{\rm m}=\pm\frac{5}{2}\right\rangle$). Error bars are smaller than the size of symbols.}
\end{figure}

To verify the preparation of a single D$_{5/2}$ Zeeman sub-level, we change the polarization of the trigger laser by rotating the quarter-wave plate in front of the \mbox{HALO} through which the light is focused onto the ion. As the \mbox{HALO} optical axis is also the quantization axis, we smoothly change between driving $\sigma^+$ and $\sigma^-$ transitions (and superpositions of the two) in the ion. Fig. \ref{fig:depopulation} shows the rate at which the D$_{5/2}$ state is depopulated during the photon generation phase as a function of the quarter-wave-plate angle. The clear sinusoidal behavior with a visibility compatible with unity confirms that only one specific $\sigma$ transition can be driven. This would still be the case if the ion were in one of the two outer Zeeman states of D$_{5/2}$, $\left|{\rm m}=+\frac{5}{2}\right\rangle$ or $\left|{\rm m}=+\frac{3}{2}\right\rangle$. However, considering also the geometry of the pump beam from $45^\circ$ during the state preparation, we conclude that only one specific D$_{5/2}$ Zeeman state ($\left|{\rm m}=+\frac{5}{2}\right\rangle$) is populated.

In order to analyze the polarization of the emitted single photons, we place a 393~nm quarter-wave plate behind one of the \mbox{HALOs}, followed by a polarizing beam-splitter cube and a PMT. The numbers of photon detection events for different wave-plate angles are shown in fig. \ref{fig:polarization}. The two data sets correspond to the two possible $\left|{\rm D}_{5/2}{\rm, m}=\pm\frac{5}{2}\right\rangle$ states that we prepare. A mean visibility of $V=90.5(6)\%$ indicates an almost pure circular photon polarization, emitted through the transitions
$\left|{\rm D}_{5/2}{\rm, m}=\pm\frac{5}{2}\right\rangle\rightarrow\left|{\rm P}_{3/2}{\rm, m}=\pm\frac{3}{2}\right\rangle\rightarrow\left|{\rm S}_{1/2}{\rm, m}=\pm\frac{1}{2}\right\rangle$, respectively. The non-unity visibility is caused by the non-perfect 393~nm quarter-wave plate. The pure $\left|\sigma\right\rangle$ polarization is corroborated by an increase of the total detection efficiency from 1.18\% to 1.55\%, as light from $\sigma$ transitions is preferentially emitted along the quantization axis (which coincides with the optical axis). Thus we have verified that we generated single photons in a pure polarization and frequency state.

In conclusion, we generated single photons at a high rate with a very small multi-photon contribution. The temporal shape of the photonic wave packet was tuned over 2 orders of magnitude by appropriate laser parameters. Importantly, in the context of quantum networks, the ion was made to emit photons of single frequency and polarization. For the future, we aim at realizing a quantum memory \cite{quantummemory} that stores the polarization of a photon absorbed on the 854~nm transition in the S$_{1/2}$ Zeeman sub-states. One possible scheme is to first prepare the ion in a coherent superposition of the two $\left|{\rm D}_{5/2}{\rm, m}=\pm\frac{5}{2}\right\rangle$ states. The absorption of an 854~nm photon will then transfer the photonic state into the ion's ground state. The absorption is heralded by the detection of the emitted 393~nm photon after projecting it onto a linear polarization basis. As shown, this heralding process is successful with up to 1.55\% probability.

\begin{acknowledgments}
We acknowledge support by the BMBF (QuOReP project, QSCALE Chist-ERA project), the German Scholars Organization / Alfried Krupp von Bohlen und Halbach-Stiftung, the EU (AQUTE Integrating Project), and the ESF (IOTA COST Action).
\end{acknowledgments}

\bibliography{bibliography}

\end{document}